\documentclass[prb,showpacsb,showpacs,preprintnumbers,twocolumn]{revtex4}
\usepackage{graphicx}
\usepackage{dcolumn}
\usepackage{bm}
\usepackage{mathrsfs}
\usepackage{graphicx}
\usepackage{dcolumn}
\usepackage{bm}

\begin{document}

\title{Spin transport properties of a quantum dot coupled to ferromagnetic
leads with noncollinear magnetizations}
\author{Hao Zhang$^{1}$, Guang-Ming Zhang$^{2}$, and Lu Yu$^{3,1}$}
\affiliation{$^1$Institute of Theoretical Physics, Chinese Academy of Sciences, 100190
Beijing, China;\\
$^{2}$Department of Physics, Tsinghua University, 100084 Beijing, China;\\
$^{3}$Institute of Physics, Chinese Academy of Sciences, 100190 Beijing,
China.}
\date{\today}

\begin{abstract}
A correct general  formula for the spin current through an
interacting quantum dot coupled to ferromagnetic leads with
magnetization at an arbitrary angle $\theta$ is derived within the
framework of the Keldysh formalism. Under asymmetric conditions, the
spin current component $J_{z}$ may change sign for $0<\theta <\pi$.
It is shown that the spin current and spin tunneling
magnetoresistance exhibit different angle dependence in the free and
Coulomb blockade regimes. In the latter case, the competition of the
spin precession and the spin-valve effect could lead to an anomaly
in the angle dependence of the spin current.
\end{abstract}

\maketitle

\section{Introduction}

In the new field of spintronics \cite{zutic}, the magnetic properties of
quantum devices control the transport properties via the electron spin, for
example, the tunnel magnetoresistance (TMR) in ferromagnetic tunnel
junctions. The high magnetoresistance in a TMR device is due to the
spin-valve effect, namely, the resistance strongly depends on whether the
magnetization of the two ferromagnetic electrodes are parallel or
antiparallel. By switching the magnetization of one electrode with respect
to the other, the charge current is modulated by the relative angle $\theta$
of the two magnetic moments. With the magnetic tunneling injection
technique, a pure spin current can be generated and detected experimentally. %
\cite{kato,valenzuela} This substantial progress in experiment makes
it feasible to investigate the spin transport properties in
mesoscopic systems.

To study the spin-dependent transport properties, a device setup of a
quantum dot (QD) coupled to ferromagnetic leads has been proposed. \cite%
{sergueev-2002} In such a geometry, the charge current can be spin
polarized and can induce a pure spin current. However, up to now,
most of the previous works were devoted to the charge transport
properties, not to the study of the spin current itself. Moreover,
the main focus was on the charge transport on a QD coupled to two
ferromagnetic leads with
collinear magnetizations, \cite%
{bulka-2003,martinek-2003,sindle-2003,choi-2003,utsumi-2005,zhang-2003,
lpez-2003,borda-2003,souza,cottet-2006,pasupathy,hamaya-2007} while
less
attention was given to the noncollinear alignment. \cite%
{konig-2003,lui,weymann,Rud-2005,swirk,braun,eto,fransson,fransson-2005,zhou,simon-2007}%
Braun \textit{et al.} \cite{braun-2005} gave an expression for the
spin current through the left tunnel barrier. However, they did not
derive an adequate unified formula for the spin current through the
two tunnel barriers and did not actually consider the spin current
in a steady state. The $z$-component of the spin current defined as
a difference between the spin-up and spin-down contributions to the
charge current was considered by Mu \textit{et al.}\cite{mu-2006}
for the noncollinear case. Unfortunately, these authors did not
properly take into account the difference of the spin quantization
axis for the two leads, so their result is correct only for the
parallel case.

Recently, Rudzi\'{n}ski \textit{et al.} \cite{Rud-2005} studied the
charge current through a quantum dot coupled to noncollinearly
polarized ferromagnetic leads. They found that the current-voltage
curve reveals typical step-like characteristics. They also found
that the spin precession is enhanced by the Coulomb correlations and
strong spin polarization of the leads. Moreover, the relationship
between the charge current and the angle of the magnetization
configurations of the electrodes has been studied by Zhou \textit{et
al.} \cite{zhou} These authors concluded that the angle dependence
of the electric current in the free regime varies monotonically from
the parallel to antiparallel alignment, while in the Coulomb
blockade regime it varies nonmonotonically. However, authors of both
references did not consider the spin current in this general
configuration.
\begin{figure}[tbph]
\centering
\par
\begin{center}
$%
\begin{array}{c@{\hspace{0.01in}}c@{\hspace{0.01in}}c}
\includegraphics[scale=0.5]{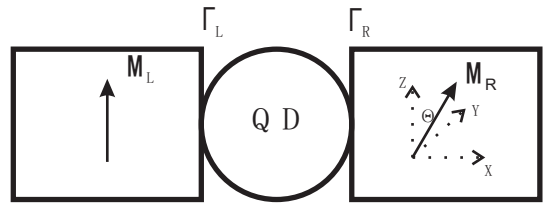} &  &  \\
&  &
\end{array}
$%
\end{center}
\caption{Sketch of the system configuration. QD is coupled to two
ferromagnetic leads with magnetizations $M_{L}$ and $M_{R}$ at an angle $%
\protect\theta $. }
\end{figure}

In this paper, we first derive an exact general formula for the spin
current through a QD coupled to noncollinear ferromagnetic leads,
starting from the Heisenberg equation for the spin operator in terms
of the Keldysh Green's functions \cite{braun-2005} (Sec.II). To the
best of our knowledge, this general formula of the spin current for
this class of devices is derived for the first time. It
should play a similar role as its charge counterpart derived earlier. \cite%
{sergueev-2002,meir-1992,jauho} Then, by using the
equation-of-motion technique with the Hartree-Fock decoupling
scheme, the spin current is obtained as a function of the bias
voltage and the angle $\theta $ of the magnetization configurations
of the leads (Sec.III). Furthermore, the spin current and the spin
tunneling magnetoresistance (STMR) are calculated numerically in
both free and Coulomb blockade regimes (Sec. IV). The interplay of
the spin precession enhanced by the Coulomb repulsion and the spin
valve effect gives rise to anomalous behavior of the angular
dependence of the spin current anticipated in the Coulomb blockade
regime.

\section{General expression for the spin current}

The system considered in this paper is schematically shown in Fig.1, and it
consists of a single-level quantum dot coupled to two ferromagnetic metallic
leads by tunneling barriers. The magnetic moment $M$ of the left electrode
is pointing to the $z$-direction, while the moment of the right electrode is
at an angle $\theta $ to the $z$ axis in the $x$-$z$ plane. We will use the
local and global quantization axes to describe the electron spin. The local
quantization axes are determined by the local spin polarization in the
leads, while the global axes are the local basis in the left electrode. The
corresponding model Hamiltonian is given by \cite{sergueev-2002}
\begin{widetext}
\begin{eqnarray}
H &=&\sum_{\mathbf{k},\sigma ;\alpha =L,R}\epsilon
_{\mathbf{k,\sigma ,\alpha }}c_{\mathbf{k},\sigma ,\alpha
}^{\dagger }c_{\mathbf{k},\sigma ,\alpha }+\sum_{\gamma }\epsilon
_{d}d_{\gamma }^{\dagger }d_{\gamma }+Ud_{\uparrow }^{\dag
}d_{\uparrow }d_{\downarrow }^{\dag }d_{\downarrow }
+\sum_{\mathbf{k}}\left[ T_{\mathbf{k,}L}\left(
c_{\mathbf{k,+,}L}^{\dag }d_{\uparrow }+c_{\mathbf{k,-,}L}^{\dag
}d_{\downarrow }\right) +h.c.\right]
\nonumber \\
&&+\sum_{\mathbf{k}}\left[ T_{\mathbf{k,}R}\left( c_{\mathbf{k,+,}%
R}^{\dagger }\cos \frac{\theta }{2}-c_{\mathbf{k,-,}R}^{\dagger }\sin \frac{%
\theta }{2}\right) d_{\uparrow }+T_{\mathbf{k,}R}\left(
c_{\mathbf{k,-,}R}^{\dagger }\cos \frac{\theta
}{2}+c_{\mathbf{k,+,}R}^{\dagger }\sin \frac{\theta }{2}\right)
d_{\downarrow }+h.c.\right],
\end{eqnarray}
where the spin projection on the local axes is denoted as $\sigma =\pm $, $%
\epsilon _{\mathbf{k,\sigma ,\alpha }}=$ $\epsilon _{\mathbf{k,\alpha }%
}+\sigma M_{\mathbf{\alpha }}$ is the single-electron energy in the $\alpha $%
-th electrode, and $c_{\mathbf{k},\sigma ,\alpha }^{\dagger }$ and $c_{%
\mathbf{k},\sigma ,\alpha }$ correspond to the creation and annihilation
operators, respectively. Similarly, the spin projection on the global axes
is denoted as $\gamma =\uparrow \downarrow $, $d_{\gamma }^{\dagger }$ and $%
d_{\gamma }$ are the creation and annihilation operators of the electron on
the quantum dot with energy $\epsilon _{d}$.

For simplicity, we can rewrite the model Hamiltonian into a
compact matrix form
\begin{equation}
H =\sum_{\mathbf{k},\alpha =L,R}\mathbf{\hat{C}}_{\mathbf{k,\alpha }%
}^{\dagger }\mathbf{\hat{\epsilon}}_{\mathbf{k,}{\alpha }}\mathbf{\hat{C}}_{%
\mathbf{k},\alpha }+\mathbf{\hat{\epsilon}}_{d}\mathbf{\hat{D}}^{\dagger }%
\mathbf{\hat{D}}
+\frac{U}{4}\left[ \left( \mathbf{\hat{D}}^{\dagger }\mathbf{\hat{D}}%
\right) ^{2}-\left( \mathbf{\hat{D}}^{\dagger }\mathbf{\hat{\sigma}}_{z}%
\mathbf{\hat{D}}\right) ^{2}\right]+\sum_{\mathbf{k},\alpha
=L,R}\left( \mathbf{\hat{C}}_{\mathbf{k},\alpha }^{\dagger
}\mathbf{\hat{T}}_{\mathbf{k,}{\alpha }}\mathbf{R}_{\alpha
}^{\dagger }\mathbf{\hat{D}+}h.c\mathbf{.}\right) ,
\end{equation}
where we have introduced the Nambu spinors and two useful matrices
\begin{equation}
\mathbf{\hat{C}_{k,\alpha}}=\left(
\begin{array}{c}
C_{\mathbf{k},+,\alpha } \\
C_{\mathbf{k},-,\alpha }%
\end{array}
\right) , \text{ }\mathbf{\hat{D}} \mathbf{=}\left(
\begin{array}{c}
d_{\uparrow } \\
d_{\downarrow }%
\end{array}%
\right) ,
\text{ }\mathbf{\hat{\epsilon}}_{\mathbf{k,}{\alpha
}}=\left(
\begin{array}{cc}
\epsilon _{\mathbf{k,+,\alpha }} & 0 \\
0 & \epsilon _{\mathbf{k,-,\alpha }}%
\end{array}%
\right) ,
\text{ \ }\mathbf{R}_{\alpha }=\left(
\begin{array}{cc}
\cos \frac{\theta _{\alpha }}{2} & -\sin \frac{\theta _{\alpha }}{2} \\
\sin \frac{\theta _{\alpha }}{2} & \cos \frac{\theta _{\alpha }}{2}%
\end{array}
\right) ,
\end{equation}
with $\theta _{L}=0$ for the left lead and $\theta _{R}=\theta $
for the right lead. When the spin operators of the two leads are
considered
$\mathbf{\hat{S}}_{\alpha }=\left( \hbar /2\right) \sum_{%
\mathbf{k}}\mathbf{\hat{C}}_{\mathbf{k},\alpha }^{\dagger }\mathbf{\hat{%
\sigma}}_{\alpha }\mathbf{\hat{C}}_{\mathbf{k},\alpha }$, the spin
matrices are given by
\begin{eqnarray}
&&\text{ }\mathbf{\sigma }_{L}^{x} =\left(
\begin{array}{cc}
0 & 1 \\
1 & 0%
\end{array}%
\right) ,\text{ \ }\mathbf{\sigma }_{L}^{y}=\left(
\begin{array}{cc}
0 & -i \\
i & 0%
\end{array}%
\right) ,  \mathbf{\sigma }_{L}^{z} =\left(
\begin{array}{cc}
1 & 0 \\
0 & -1%
\end{array}%
\right) , \nonumber \\
&&\mathbf{\sigma }_{R}^{x}=\left(
\begin{array}{cc}
\sin \theta & \cos \theta \\
\cos \theta & -\sin \theta%
\end{array}%
\right) , \mathbf{\sigma }_{R}^{y} =\mathbf{\sigma }_{L}^{y},
\mathbf{\sigma }_{R}^{z}=\left(
\begin{array}{cc}
\cos \theta & -\sin \theta \\
-\sin \theta & -\cos \theta%
\end{array}%
\right) .
\end{eqnarray}
From the Heisenberg equation, we can calculate the spin current
$\mathbf{J}_{\alpha }=\langle \mathbf{\hat{J}}_{\alpha }\rangle $
from the lead to the dot\cite{braun-2005}
\begin{equation}
\mathbf{\hat{J}}_{\alpha } =\frac{i}{\hbar
}[\mathbf{\hat{S}}_{\alpha },H]
=\frac{i}{2}\sum_{\mathbf{k}}\mathrm{Tr}\left( \mathbf{\hat{C}}_{\mathbf{k}%
,\alpha }^{\dagger }\mathbf{\hat{\sigma}}_{\alpha }\mathbf{\hat{T}}_{\mathbf{%
k,}{\alpha }}\mathbf{R}_{\alpha }^{\dagger }\mathbf{\hat{D}}-\mathbf{\hat{D}}%
^{\dagger }\mathbf{R}_{\alpha
}\mathbf{\hat{T}}_{\mathbf{k,}{\alpha }}^{\ast
}\mathbf{\hat{\sigma}}_{\alpha
}\mathbf{\hat{C}}_{\mathbf{k},\alpha }\right).
\end{equation}
Moreover, by introducing the Keldysh Green's function matrices
\begin{equation}
\mathbf{\hat{G}}_{\alpha }^{<}\left( \mathbf{k},t\right) =i\left(
\begin{array}{cc}
\langle C_{\mathbf{k},+,\alpha }^{\dagger }\left( 0\right) d_{\uparrow
}\left( t\right) \rangle & \langle C_{\mathbf{k},-,\alpha }^{\dagger }\left(
0\right) d_{\uparrow }\left( t\right) \rangle \\
\langle C_{\mathbf{k},+,\alpha }^{\dagger }\left( 0\right) d_{\downarrow
}\left( t\right) \rangle & \langle C_{\mathbf{k},-,\alpha }^{\dagger }\left(
0\right) d_{\downarrow }\left( t\right) \rangle%
\end{array}%
\right) ,  \mathbf{\hat{G}}_{d}^{<}\left( t\right) = i\left(
\begin{array}{cc}
\langle d_{\mathbf{\uparrow }}^{\dagger }\left( 0\right) d_{\uparrow }\left(
t\right) \rangle & \langle d_{\mathbf{\downarrow }}^{\dagger }\left(
0\right) d_{\uparrow }\left( t\right) \rangle \\
\langle d_{\mathbf{\uparrow }}^{\dagger }\left( 0\right) d_{\downarrow
}\left( t\right) \rangle & \langle d_{\mathbf{\downarrow }}^{\dagger }\left(
0\right) d_{\downarrow }\left( t\right) \rangle%
\end{array}%
\right) ,
\end{equation}
we can further rewrite the expectation value of the spin current
as
\begin{equation}
\mathbf{J}_{\alpha }=\sum_{\mathbf{k}}\int \frac{d\omega }{2\pi }\mathrm{Re}%
\left[ \mathrm{Tr}\left( \mathbf{\hat{G}}_{\mathbf{\alpha }}^{<}\left(
\mathbf{k},\omega \right) \mathbf{\hat{\sigma}}_{\alpha }\mathbf{\hat{T}}_{%
\mathbf{k,}{\alpha }}\mathbf{R}_{\alpha }^{\dagger }\right) \right] ,
\end{equation}%
where $\mathbf{\hat{G}}_{\mathbf{\alpha }}^{<}\left( \mathbf{k},\omega
\right) $ is the Fourier transform of $\mathbf{\hat{G}}_{\alpha }^{<}\left(
\mathbf{k},t\right) $. Since the ferromagnetic leads are noninteracting, we
obtain the Dyson equation for $\mathbf{\hat{G}}_{\mathbf{\alpha }}^{<}\left(
\mathbf{k},\omega \right) $ in terms of the Green's function matrices for
the local dot electrons,
\begin{equation}
\mathbf{\hat{G}}_{\mathbf{\alpha }}^{<}\left( \mathbf{k},\omega
\right) =
\mathbf{\hat{G}}_{\mathbf{d}}^{\mathbf{r}}\left( \omega \right) \mathbf{R}%
_{\alpha }\mathbf{\hat{T}}_{\mathbf{k,}{\alpha }}^{\ast }\mathbf{\hat{g}}%
_{\alpha }^{<}\left( \mathbf{k,}\omega \right)
+\mathbf{\hat{G}}_{\mathbf{d}}^{\mathbf{<}}\left( \omega \right) \mathbf{R}%
_{\alpha }\mathbf{\hat{T}}_{\mathbf{k,}{\alpha }}^{\ast }\mathbf{\hat{g}}%
_{\alpha }^{\mathbf{a}}\left( \mathbf{k,}\omega \right) ,
\end{equation}
where
\begin{eqnarray*}
\mathbf{\hat{g}}_{\alpha }^{<}\left( \mathbf{k,}\omega \right) &=&2\pi
if_{\alpha }\left( \omega \right) \left(
\begin{array}{cc}
\delta \left( \omega -\epsilon _{\mathbf{k,+,\alpha }}\right) & 0 \\
0 & \delta \left( \omega -\epsilon _{\mathbf{k,-,\alpha }}\right)%
\end{array}%
\right) , \\
\mathbf{\hat{g}}_{\alpha }^{\mathbf{a}}\left( \mathbf{k,}\omega \right)
&=&\left(
\begin{array}{cc}
\frac{1}{\omega -\epsilon _{\mathbf{k,+,\alpha }}-i0^{+}} & 0 \\
0 & \frac{1}{\omega -\epsilon _{\mathbf{k,-,\alpha }}-i0^{+}}%
\end{array}
\right),
\end{eqnarray*}
with $f_{\alpha}(\omega) = [1+\exp(\omega -\mu
_{\alpha})/(k_BT)]^{-1}$, $\mu _{L}=-eV/2$ and $\mu _{R}=eV/2$.
Inserting these expressions into the spin current formula, we
obtain the spin current as follows:
\begin{equation}
\mathbf{J}_{\alpha }=\int \frac{d\omega }{4\pi }\mathrm{Re}\left( \mathrm{Tr}%
\left\{ i\text{ }\mathbf{\tilde{\Gamma}}_{\alpha }\left( \omega \right) %
\left[ 2f_{\alpha }\left( \omega \right) \mathbf{\hat{G}}_{\mathbf{d}}^{%
\mathbf{r}}\left( \omega \right) +\mathbf{\hat{G}}_{\mathbf{d}}^{\mathbf{<}%
}\left( \omega \right) -i\mathcal{P}\int \frac{dE}{\pi }\frac{\mathbf{\hat{G}%
}_{\mathbf{d}}^{\mathbf{<}}\left( E\right) }{E-\omega }\right] \right\}
\right) ,
\end{equation}%
where the integral is taken as the principle value and%
\begin{equation}
\mathbf{\tilde{\Gamma}}_{\alpha }\left( \omega \right) =\mathbf{R}_{\alpha
}\left(
\begin{array}{cc}
\Gamma _{+,\alpha }\left( \omega \right)  & 0 \\
0 & \Gamma _{-,\alpha }\left( \omega \right)
\end{array}%
\right) \mathbf{\hat{\sigma}}_{\alpha }\mathbf{R}_{\alpha }^{\dagger },\text{
}\Gamma _{\sigma ,\alpha }\left( \omega \right) =2\pi \sum_{\mathbf{k}%
}\left| T_{\mathbf{k,}{\alpha }}\right| ^{2}\delta \left( \omega -\epsilon _{%
\mathbf{k,\sigma ,\alpha }}\right) .
\end{equation}%
Since this system is quasi one-dimensional, different from the
spin Hall systems\cite{mura,sino,bern} in which the spin-orbit
coupling is essential, we do not take into account those spin flip
processes due to the spin-orbit coupling. So we do consider the
spin current through QD as a continuous and conserved quantity.
The steady state is realized in the system through the scattering
process which is similar to the charge transport. As far as we
understand, no one has studied the detailed relaxation process
within the QD.
In a steady state, the spin current is uniform, so $\mathbf{J}_{L}=-%
\mathbf{J}_{R}$. Thus, we can symmetrize the spin current as $\mathbf{J}%
=(\mathbf{J}_{L}-\mathbf{J}_{R})/2$ which is similar to the
operation performed on the expression for the charge current
\cite{sergueev-2002,meir-1992,jauho}. The general expression for
the spin
current is then given by%
\begin{equation}
\mathbf{J}\mathbf{=}\frac{1}{2}\int \frac{d\omega }{2\pi }\mathrm{Re}\left\{
\mathrm{Tr}\left[ i\left[ f_{L}\left( \omega \right) \mathbf{\tilde{\Gamma}}%
_{L}\left( \omega \right) -f_{R}\left( \omega \right) \mathbf{\tilde{\Gamma}}%
_{R}\left( \omega \right) \right] \mathbf{\hat{G}}_{\mathbf{d}}^{\mathbf{r}%
}\left( \omega \right) +\left[ \mathbf{\tilde{\Gamma}}_{L}\left( \omega
\right) -\mathbf{\tilde{\Gamma}}_{R}\left( \omega \right) \right] \left(
\frac{i}{2}\mathbf{\hat{G}}_{\mathbf{d}}^{\mathbf{<}}\left( \omega \right) +%
\mathcal{P}\int \frac{dE}{2\pi }\frac{\mathbf{\hat{G}}_{\mathbf{d}}^{\mathbf{%
<}}\left( E\right) }{E-\omega }\right) \right] \right\} .
\end{equation}
\end{widetext}
Braun \textit{et al.} \cite{braun-2005} gave an expression for the
spin current through the left tunnel barrier, but they did not
derive an unified formula for the spin current through the left
and right tunnel barriers. Also, these authors did not provide a
symmetrized formula in the steady state, which is essential for
the calculation and discussion of the spin current. Mu \textit{et
al.} \cite{mu-2006} used the difference between the charge
currents through the spin up and down channels to define the
$z$-component of the spin current. However, these authors did not
properly take into account the difference of the two local
quantization axes of the two ferromagnetic leads which strongly
affects the tunneling hamiltonian as pointed out by Rudzi\'{n}ski
\textit{et al.}\cite{Rud-2005} Moreover, their expression of the
charge current was not symmetrized. As a result, their formula is
correct only for the parallel case.

\section{Calculation of the Keldysh Green's functions}

To investigate the nonequilibrium transport properties, there are two
commonly used techniques to calculate the Keldysh Green's functions. One is
the real-time diagrammatic technique, \cite%
{schoeller,konig,utsumi-2005,weymann,braun} based on a perturbation
expansion in terms of the dot-lead coupling strength, whereas the Coulomb
interactions on the dot are exactly taken into account. However, this
technique only considers finite order tunneling processes, and cannot deal
with the coupling between the dot and the electrode exactly. The other
alternative is the equation-of-motion technique\cite%
{haug,sergueev-2002,zhang-2003,souza,Rud-2005,swirk} which treats the
dot-lead coupling exactly, while the strong correlations on the dot can be
dealt with only approximately.

In this paper, the Green's functions are solved by the
equation-of-motion technique with the Hartree-Fock decoupling
scheme.\cite{Rud-2005,haug} The solution can be written in a compact form of the matrix Dyson equation%
\begin{equation}
\mathbf{\hat{G}}_{\mathbf{d}}\left( \omega \right) =\left[ \mathbf{1-\hat{g}}%
_{\mathbf{d}}\left( \omega \right) \mathbf{\Sigma }^{\left( 0\right) }\left(
\omega \right) \right] ^{-1}\mathbf{\hat{g}}_{\mathbf{d}}\left( \omega
\right) ,
\end{equation}%
where%
\[
\mathbf{\hat{g}}_{\mathbf{d}}\left( \omega \right) =\left(
\begin{array}{cc}
\frac{\omega -\epsilon _{d}-U\left( 1-\left\langle n_{\downarrow ,\downarrow
}\right\rangle \right) }{\left( \omega -\epsilon _{d}\right) \left( \omega
-\epsilon _{d}-U\right) } & -\frac{U\left\langle n_{\downarrow ,\uparrow
}\right\rangle }{\left( \omega -\epsilon _{d}\right) \left( \omega -\epsilon
_{d}-U\right) } \\
-\frac{U\left\langle n_{\uparrow ,\downarrow }\right\rangle }{\left( \omega
-\epsilon _{d}\right) \left( \omega -\epsilon _{d}-U\right) } & \frac{\omega
-\epsilon _{d}-U\left( 1-\left\langle n_{\uparrow ,\uparrow }\right\rangle
\right) }{\left( \omega -\epsilon _{d}\right) \left( \omega -\epsilon
_{d}-U\right) }%
\end{array}%
\right)
\]%
with $\left\langle n_{\alpha ,\beta }\right\rangle =\left\langle d_{\alpha
}^{\dagger }d_{\beta }\right\rangle $ and $\mathbf{\hat{g}}_{\mathbf{d}%
}\left( \omega \right) $ denotes the corresponding Green functions in the
matrix form of the uncoupled dot. The self-energy $\mathbf{\Sigma }^{\left(
0\right) }\left( \omega \right) $ is given by%
\[
\mathbf{\Sigma }^{\left( 0\right) }\left( \omega \right) =\left(
\begin{array}{cc}
\Sigma _{++}^{\left( 0\right) }\left( \omega \right)  & \Sigma _{+-}^{\left(
0\right) }\left( \omega \right)  \\
\Sigma _{-+}^{\left( 0\right) }\left( \omega \right)  & \Sigma _{--}^{\left(
0\right) }\left( \omega \right)
\end{array}%
\right) ,
\]%
with%
\begin{eqnarray}
\Sigma _{\pm \pm }^{\left( 0\right) }\left( \omega \right)  &=&\sum_{\mathbf{%
k}}\left[ \frac{|T_{\mathbf{k,}\emph{L}}|^{2}}{\omega -\epsilon _{\mathbf{%
k,\pm ,}L}}+C_{\mathbf{K}}\left| T_{\mathbf{k,}\emph{R}}\right| ^{2}\right] ,
\nonumber \\
\Sigma _{\pm \mp }^{\left( 0\right) }\left( \omega \right)  &=&\frac{1}{2}%
\sum_{\mathbf{k}}\left| T_{\mathbf{k,}\emph{R}}\right| ^{2}D_{\mathbf{K}%
}\sin \theta ,  \nonumber \\
C_{\mathbf{K}} &=&\frac{\cos ^{2}\left( \theta /2\right) }{\omega -\epsilon
_{\mathbf{k,\pm ,}R}}+\frac{\sin ^{2}\left( \theta /2\right) }{\omega
-\epsilon _{\mathbf{k,\mp ,}R}},  \nonumber \\
D_{\mathbf{K}} &=&\frac{1}{\omega -\epsilon _{\mathbf{k,+,}R}}-\frac{1}{%
\omega -\epsilon _{\mathbf{k,-,}R}}.
\end{eqnarray}%
Then one can calculate the retarded Green functions as%
\[
G_{\uparrow \uparrow }^{r}\left( \omega \right) =\left[ g_{\uparrow \uparrow
}^{r}\left( \omega \right) -A(\omega )\Sigma _{--}^{\left( 0\right) r}\left(
\omega \right) \right] /B(\omega ),
\]%
\[
G_{\uparrow \downarrow }^{r}\left( \omega \right) =\left[ g_{\uparrow
\downarrow }^{r}\left( \omega \right) -A(\omega )\Sigma _{+-}^{\left(
0\right) r}\left( \omega \right) \right] /B(\omega ),
\]%
\[
G_{\downarrow \uparrow }^{r}\left( \omega \right) =\left[ g_{\downarrow
\uparrow }^{r}\left( \omega \right) -A(\omega )\Sigma _{-+}^{\left( 0\right)
r}\left( \omega \right) \right] /B(\omega ),
\]%
\[
G_{\downarrow \downarrow }^{r}\left( \omega \right) =\left[ g_{\downarrow
\downarrow }^{r}\left( \omega \right) -A(\omega )\Sigma _{++}^{\left(
0\right) r}\left( \omega \right) \right] /B(\omega ),
\]%
where%
\begin{eqnarray*}
A(\omega ) &=&g_{\uparrow \uparrow }^{r}\left( \omega \right) g_{\downarrow
\downarrow }^{r}\left( \omega \right) -g_{\uparrow \downarrow }^{r}\left(
\omega \right) g_{\downarrow \uparrow }^{r}\left( \omega \right) , \\
B(\omega ) &=&1-g_{\uparrow \uparrow }^{r}\left( \omega \right) \Sigma
_{++}^{\left( 0\right) r}\left( \omega \right) -g_{\downarrow \downarrow
}^{r}\left( \omega \right) \Sigma _{--}^{\left( 0\right) r}\left( \omega
\right)  \\
&&-g_{\uparrow \downarrow }^{r}\left( \omega \right) \Sigma _{-+}^{\left(
0\right) r}\left( \omega \right) -g_{\downarrow \uparrow }^{r}\left( \omega
\right) \Sigma _{+-}^{\left( 0\right) r}\left( \omega \right)  \\
&&+A(\omega )\left[ \Sigma _{++}^{\left( 0\right) r}\left( \omega \right)
\Sigma _{--}^{\left( 0\right) r}\left( \omega \right) -\Sigma _{+-}^{\left(
0\right) r}\left( \omega \right) \Sigma _{-+}^{\left( 0\right) r}\left(
\omega \right) \right] .
\end{eqnarray*}%
The retarded self-energies $\Sigma _{\pm \pm }^{\left( 0\right) r}\left(
\omega \right) $ and $\Sigma _{\pm \mp }^{\left( 0\right) r}\left( \omega
\right) $ are given by the formulas%
\begin{eqnarray}
\Sigma _{\pm \pm }^{\left( 0\right) r}\left( \omega \right)  &=&-\frac{i}{2}%
[\Gamma _{\pm ,L}\left( \omega \right) +\Gamma _{\pm ,R}(\omega )\cos
^{2}\left( \theta /2\right)   \nonumber \\
&&+\Gamma _{\mp ,R}(\omega )\sin ^{2}\left( \theta /2\right) ],  \nonumber \\
\Sigma _{\pm \mp }^{\left( 0\right) r}\left( \omega \right)  &=&-\frac{i}{4}%
\left[ \Gamma _{+,R}(\omega )-\Gamma _{-,R}(\omega )\right] \sin \theta .
\end{eqnarray}

In the following we assume%
\begin{eqnarray*}
\Gamma _{\pm ,L}\left( \omega \right) &=&\Gamma _{\pm ,L}=\Gamma _{0}\left(
1\pm p_{l}\right) , \\
\Gamma _{\pm ,R}(\omega ) &=&\Gamma _{\pm ,R}=\gamma \Gamma _{0}\left( 1\pm
p_{r}\right) ,
\end{eqnarray*}%
where $p_{l}$ and $p_{r}$ denote the spin polarization of the left and right
electrodes, respectively, and the parameter $\gamma $ expresses the
asymmetry coupling of the left and right electrodes to the dot. $\mathbf{%
\hat{G}}_{\mathbf{d}}^{\mathbf{<}}\left( \omega \right) $ can be obtained
from the Keldysh equation,%
\begin{equation}
\mathbf{\hat{G}}_{\mathbf{d}}^{\mathbf{<}}\left( \omega \right) =\mathbf{%
\hat{G}}_{\mathbf{d}}^{\mathbf{r}}\left( \omega \right) \mathbf{\Sigma }%
^{<}\left( \omega \right) \mathbf{\hat{G}}_{\mathbf{d}}^{\mathbf{a}}\left(
\omega \right) ,
\end{equation}%
where the full self-energy $\mathbf{\Sigma }^{<}\left( \omega \right) $ is
related to $\mathbf{\Sigma }^{(0)<}\left( \omega \right) $ via the Ng ansatz %
\cite{Ng}%
\begin{eqnarray}
\Sigma _{\pm \pm }^{<}\left( \omega \right) &=&i\Gamma _{0}[f_{L}\left(
\omega \right) \left( 1\pm p_{l}\right) +\gamma f_{R}\left( \omega \right)
\left( 1\pm p_{r}\cos \theta \right) ],  \nonumber \\
\Sigma _{\pm \mp }^{<}\left( \omega \right) &=&i\gamma \Gamma
_{0}f_{R}\left( \omega \right) p_{r}\sin \theta .
\end{eqnarray}%
The statistical averages of $\left\langle n_{\alpha ,\beta }\right\rangle $
have to be calculated self-consistently in the following way:%
\begin{eqnarray}
\left\langle n_{\sigma ,\sigma }\right\rangle &=&\mathrm{Im}\int_{-\infty
}^{+\infty }\frac{d\omega }{2\pi }G_{\sigma \sigma }^{<}\left( \omega
\right) ,  \nonumber \\
\left\langle n_{\sigma ,\bar{\sigma}}\right\rangle &=&-i\int_{-\infty
}^{+\infty }\frac{d\omega }{2\pi }G_{\bar{\sigma}\sigma }^{<}\left( \omega
\right) .
\end{eqnarray}

This approximate calculation of the Keldysh Green's functions does
not take into account the Kondo-like correlations which need a
careful treatment of the Coulomb interaction on the dot. Some
previous works which studied the charge transport properties of
this system discussed the Kondo effect, including the collinear
alignment\cite{bulka-2003,martinek-2003,sindle-2003,choi-2003,utsumi-2005,zhang-2003,lpez-2003,borda-2003,pasupathy,hamaya-2007}
and the noncollinear
case\cite{sergueev-2002,swirk,eto,simon-2007}. It is left for our
future work to discuss the influence of the Kondo-like
correlations on the spin transport properties of this system.

\section{Results and discussions}

Now we numerically calculate the three components of the spin current. Since
a general magnetic configuration of the leads is considered, the spin
tunneling magnetoresistance (STMR) can be estimated by%
\begin{equation}
STMR_{a}=\frac{J_{a}(\theta =0)-J_{a}(\theta )}{J_{a}(\theta =0)},a=x,y,z
\end{equation}%
where $J_{x,y,z}(\theta )$ denote the three components of the spin current.
In the following three different situations are considered: a symmetric
junction with fully polarized external electrodes ($p_{l}=p_{r}=1$), with
partially polarized external electrodes ($p_{l}=p_{r}=0.4$), and an
asymmetric junction ($p_{l}=0.4,p_{r}=1$).

The $J_{z}$-voltage curve for the symmetric cases reveals typical step-like
characteristics. Below the lower threshold voltage, the dot is empty and the
sequential contribution to $J_{z}$ is exponentially suppressed. The first
step in $J_{z}$ occurs at a critical bias, where the discrete level $%
\epsilon _{d}$ crosses the Fermi level, whereas the step at a higher
threshold corresponds to the case when $\epsilon _{d}+U$ crosses the Fermi
level. In the same voltage range, $J_{z}$ in the case of $p_{l}=p_{r}=1$ is
much larger, since the external electrodes are fully polarized. Mu \textit{%
et al.} \cite{mu-2006} also considered this case ($p_{l}=p_{r}=0.4,\theta
=\pi /3$), but their result is different from ours because they did not
properly take into account the difference of the local quantization axes in
the two leads and hence the result for the $z$-component of the spin current
is incorrect. The case of $p_{l}=0.4,p_{r}=1$ is more complex (red dashed
curve in Fig.1(a)), as the asymmetry between the left and right electrodes
gives rise to asymmetrical transport characteristics of the junction with
respect to the bias reversal. For the positive bias, $J_{z}$ curve is rather
smooth above the first threshold voltage, while for the negative bias, below
the first threshold sequential tunneling is exponentially suppressed and
only the higher-order tunneling processes are possible. When $\epsilon _{d}$
approaches the Fermi level of the left electrode, the resonant tunneling can
happen. However, as the bias further increases, $J_{z}$ is suppressed by an
electron on the QD since the electrode is partially polarized (Coulomb
blockade effect), and a small peak appears as a result of competition
between the resonance tunneling and the Coulomb repulsion. After the second
resonant tunneling, $J_{z}$ finally saturates at a certain level. The
behavior of spin current component $J_{x}$ (Fig.2(b)) is similar to the
component $J_{z}$ (Fig.2(a)), because the magnetizations of the two leads
are aligned in the $x$-$z$ plane. However, the asymmetry effect resulting in
the appearance of a peak at the first threshold is more pronounced. It
appears even for the symmetric electrodes ($p_{l}=p_{r}=0.4,1$), because the
Coulomb blockade effect already shows up. The asymmetry of the spin current
curve is even more pronounced for the $y-$ component (Fig.2(c)).
Nevertheless, the two peaks on the $J_{y}$ curve are exactly located at the
two resonant tunneling biases.

\begin{figure}[tbph]
\centering
\par
\begin{center}
$%
\begin{array}{c@{\hspace{0.01in}}c@{\hspace{0.01in}}c}
\includegraphics[scale=1.2]{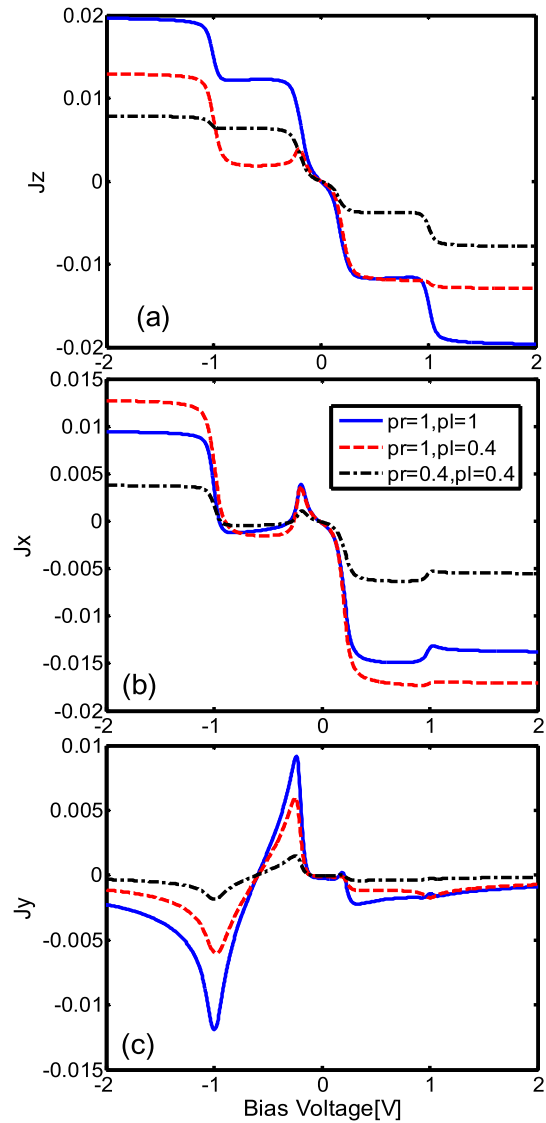} &  &  \\
&  &
\end{array}
$%
\end{center}
\caption{Voltage bias dependence of the spin current for $\protect\theta =%
\protect\pi /3$. (a) $J_{z}$, (b) $J_{x}$, and (c) $J_{y}$. The parameter
values assumed are: $\protect\epsilon _{d}=0.1eV$, $U=0.4eV$, $\Gamma
_{0}=0.01eV$, $\protect\gamma =1$, and $T=100K$. }
\end{figure}

The spin current is strongly affected by the angle $\theta $ between the
magnetic moments of the leads and we can use STMR to describe it. In the
free regime, where $\left| eV/2\right| >\epsilon _{d}+U$, the QD energy
level may be occupied by two electrons, because the Coulomb correlation
plays a little role in the spin tunneling. As a result, $J_{z}$ and $%
STMR_{z} $ exhibit a monotonic variation between the parallel and
antiparallel magnetization configurations, which is typical of a
normal spin-valve effect. Under the third condition
($p_{l}=0.4,p_{r}=1$), $J_{z}$ can achieve a negative value. The
absolute values of $J_{x}$ and $J_{y}$ achieve their maxima between
$\theta =0$ and $\theta =\pi $ as shown in Fig.3(c) and Fig.3(e),
since the absolute values of the $x$ and $y$ components of the
electron spin in the right electrode may increase when the magnetic
moment of the right lead approaches the $x$-$y$ plane.
\begin{figure}[tbph]
\centering
\par
\begin{center}
\includegraphics[scale=0.7]{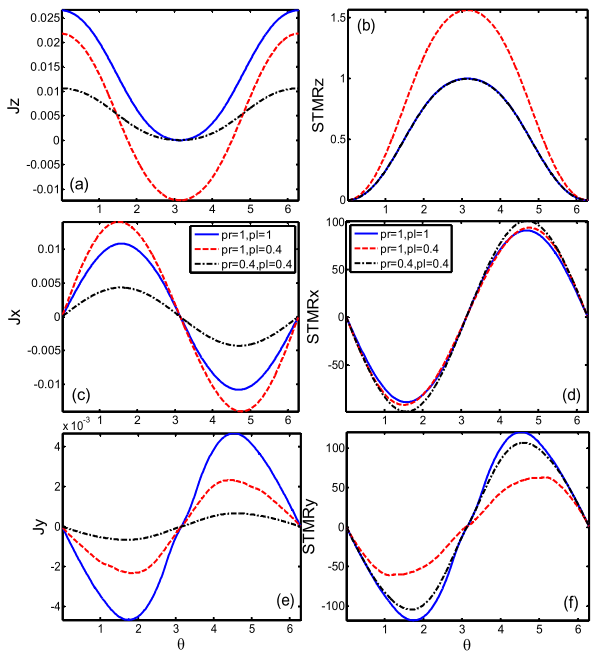}
\end{center}
\caption{Angle dependence of the spin current and spin tunneling
magnetoresistance in the free regime for v=-1.5V. (a) $J_{z}$, (b) $STMR_{z}$%
, (c) $J_{x}$, (d) $STMR_{x}$, (e) $J_{y}$, and (f) $STMR_{y}$.
The
parameter values assumed are: $\protect\epsilon _{d}=0.1eV$, $U=0.4eV$, $%
\Gamma _{0}=0.01eV$, $\protect\gamma =1$, and $T=100$K. }
\end{figure}

In the Coulomb blockade regime $\epsilon _{d}<\left| eV/2\right| <\epsilon
_{d}+U$, the QD energy level can be occupied only by one electron. The
Coulomb interaction plays an important role in the spin current through the
QD. In Fig.4(a), it is found that $J_{z}\left( \theta =0\right) $ is no
longer maximal and $J_{z}\left( \theta \right) $ is greater than $%
J_{z}\left( \theta =0\right) $ in a wide range of $\theta $ under this
condition ($p_{l}=0.4,p_{r}=1$). It is quite different from that in the free
regime. The coupling between the QD and the ferromagnetic leads may induce
an effective exchange field, and its strength and orientation with respect
to the global quantization axis depend on the bias voltage and the angle
between magnetizations of the leads. When only one electron resides on the
QD energy level, the spin degrees of freedom experience a torque due to the
effective exchange field, which results in precession of the spin around the
field. \cite{konig-2003} This process would suppress $J_{z}$, and the
competition of the spin precession effect and the spin-valve effect leads to
the anomaly of $J_{z}\left( \theta \right) $. As a result of the spin
precession, the signs of $J_{x}\left( \theta \right) $ and $J_{y}\left(
\theta \right) $ are opposite to those in the free regime.

\begin{figure}[tbph]
\centering
\par
\begin{center}
\includegraphics[scale=0.7]{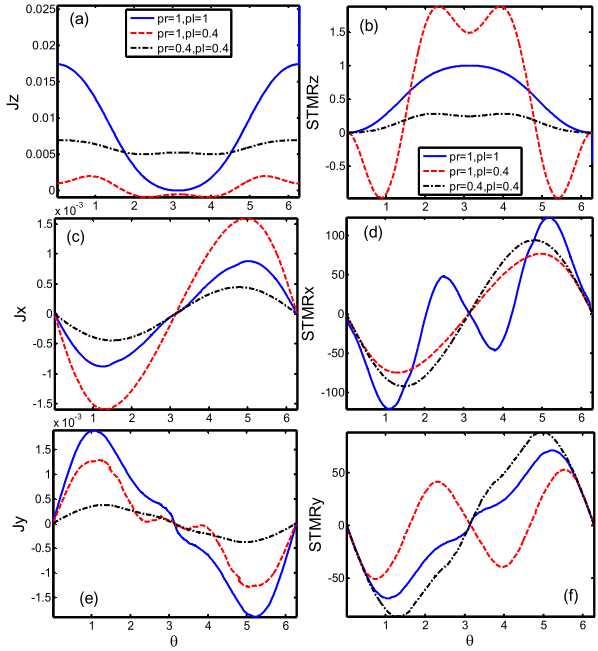}
\end{center}
\caption{Angle dependence of the spin current and spin tunneling
magnetoresistance in the Coulomb blockade regime for v=-0.5V. (a)
$J_{z}$,
(b) $STMR_{z}$, (c) $J_{x}$, (d) $STMR_{x}$, (e) $J_{y}$, and (f) $STMR_{y}$%
. The parameters assumed for numerical calculations are: $\protect\epsilon %
_{d}=0.1eV$, $U=0.4eV$, $\Gamma _{0}=0.01eV$, $\protect\gamma =1$, and $%
T=100 $K. }
\end{figure}

In conclusion, we have derived a general formula for the spin
current through a QD coupled to ferromagnetic leads with
noncollinear magnetizations, and used the formula to calculate the
spin transport properties of the system. The competition of the
spin precession and the spin-valve effect results in an anomaly of
the angle dependence of the spin current. Further investigations
are needed to carefully treat the Coulomb interaction on the QD.

\acknowledgements We would like to thank Q. F. Sun for helpful
discussions. The financial support from the Chinese Academy of
Sciences (CAS), the Ministry of Science and Technology (MOST) and
NSF-China (Grant No.10734110 ) is gratefully acknowledged.

\end{document}